\documentclass[%
 reprint,
 amsmath,amssymb,
 aps,
]{revtex4-2}
\usepackage{physunits}
\usepackage{graphicx}% Include figure files
\usepackage{dcolumn}% Align table columns on decimal point
\usepackage{bm}% bold math
\usepackage{hyperref}% add hypertext capabilities
\usepackage{siunitx}% 
\usepackage{nicefrac}%braket stuff
\newcommand{\sket}[1]{{\ensuremath{\lvert#1\rangle}}}
\newcommand{\lket}[1]{{\ensuremath{\left\lvert#1\right\rangle}}}
\newcommand{\ket}[1]{\if@display\lket{#1}\else\sket{#1}\fi}

\newcommand{\sbra}[1]{{\ensuremath{\langle#1\rvert}}}
\newcommand{\lbra}[1]{{\ensuremath{\left\langle#1\right\rvert}}}
\newcommand{\bra}[1]{\if@display\lbra{#1}\else\sbra{#1}\fi}

\newcommand{\sbraket}[2]{{\ensuremath{\langle#1\rvert#2\rangle}}}
\newcommand{\lbraket}[2]{{\ensuremath{\left\langle#1\!\left\rvert\vphantom{#1}#2\right.\!\right\rangle}}}
\newcommand{\braket}[2]{\if@display\lbraket{#1}{#2}\else\sbraket{#1}{#2}\fi}

\newcommand{\sketbra}[2]{{\ensuremath{\lvert #1\rangle\!\langle #2\rvert}}}
\newcommand{\lketbra}[2]{{\ensuremath{\left\lvert #1\right\rangle\!\!\left\langle #2\right\rvert}}}
\newcommand{\ketbra}[2]{\if@display\lketbra{#1}{#2}\else\sketbra{#1}{#2}\fi}
\usepackage{tcolorbox}
\usepackage{mathtools}

\newcommand{\proj}[1]{\ketbra{#1}{#1}}

\begin{document}

\preprint{APS/123-QED}

%\title{Experimental quantum nonlocality with AlGaAs multiplexed entangled photon source simulating triangular quantum networks}
%\title{Experimental quantum triangle network nonlocality with correlated sources}
\title{Experimental quantum triangle network nonlocality with an AlGaAs multiplexed entangled photon source}
% or with non-independent sources?

% Force line breaks with \\
%\title{Title}% Force line breaks with \\

\author{ O. Meskine$^{1}$}
\author{ I. \v{S}upi\'{c}$^{2}$}
\author{ D. Markham$^2$}
\author{ F. Appas$^3$}
\author{ F. Boitier$^{4}$}
\author{ M. Morassi$^{5}$}
\author{ A. Lemaître$^{5}$}
\author{ M.I. Amanti$^{1}$}
\author{ F. Baboux$^1$}
\author{ E. Diamanti$^2$}
\author{ S. Ducci$^1$}
%\email{corresponding autho: sara.ducci@u-paris.fr}

\affiliation{$^{1}$Université Paris Cité, CNRS UMR 7162, Laboratoire Matériaux et Phénomènes Quantiques, 75013, Paris, France}
\affiliation{$^{2}$Sorbonne Université, CNRS, LIP6, 75005 Paris, France}
\affiliation{$^{3}$ICFO-Institut de Ciencies Fotoniques, The Barcelona Institute of Science and Technology, Castelldefels (Barcelona) 08860, Spain}
\affiliation{$^{4}$Nokia Bell Labs, 91300, Massy, France}
\affiliation{$^{5}$Université Paris-Saclay, CNRS, Centre de Nanosciences et de Nanotechnologies, 91120, Palaiseau, France}

\date{\today}% It is always \today, today,
             %  but any date may be explicitly specified

\begin{abstract}
The exploration of the concept of nonlocality beyond standard Bell scenarios in quantum network architectures unveils fundamentally new forms of correlations that hold a strong potential for future applications of quantum communication networks. To materialize this potential, it is necessary to adapt theoretical advances to realistic configurations. Here we consider a quantum triangle network, for which is was shown in theory that, remarkably, quantum nonlocality without inputs can be demonstrated for sources with an arbitrarily small level of independence. We realize experimentally such correlated sources by carefully engineering the output state of a single AlGaAs multiplexed entangled-photon source, exploiting energy-matched channels cut in its broad spectrum. This simulated triangle network is then used to violate experimentally for the first time a Bell-like inequality that we derive to capture the effect of noise in the correlations present in our system. We also rigorously validate our findings by analysing the mutual information between the generated states. Our results allow us to deepen our understanding of network nonlocality while also pushing its practical relevance for quantum communication networks.
\end{abstract}
%\keywords{Suggested keywords}%Use showkeys class option if keyword
                              %display desired
\maketitle

%\tableofcontents

\section{\label{sec:level1}Introduction}

The study of quantum nonlocality within network structures has attracted increasing attention in recent years, offering a departure from traditional Bell scenarios. Unlike Bell tests, which typically involve a single source distributing entangled states to distant observers, network setups feature multiple independent sources disseminating entanglement across interconnected parties. The combination of entangled states and joint measurements within network architectures leads to correlations that are intrinsic to the network topology itself, showcasing a range of quantum nonlocal phenomena previously unexplored. 
One notable observation is the manifestation of quantum nonlocality without the requirement for distinct measurement settings, a phenomenon termed ``quantum nonlocality without inputs''~\cite{Fritz_2012,Renou_2019}. In this scenario, parties execute fixed measurements independently, yielding correlations that can defy classical explanations. 

The simplest network scenario exhibiting quantum nonlocality without inputs comprises three independent sources, each distributing particles to a pair of parties, forming a triangle structure. In this setting the three parties, usually named Alice, Bob, and Charlie use their quantum measurement to sample outputs, respectively denoted with $a$, $b$, and $c$. By collecting the outputs from a repeated experiment one can estimate the sampling probability distribution $p(a,b,c)$. Fritz showed that by using quantum resources the parties can sample a distribution that cannot be simulated with classical resources~\cite{Fritz_2012}. Classical resources in this context define a network hidden variable model, in which each pair of parties shares a hidden variable, meaning that every party has access to two hidden variables. Respecting the triangle network architecture the three hidden variables are mutually independent. Based on the values of two accessible hidden variables the players choose the output in each round of the experiment. Fritz explicitly uses the condition that the mutual information among three hidden variables is equal to zero, to show that such a model cannot reproduce statistics that the players can obtain by measuring carefully chosen quantum states. After this first example of network nonlocality, a few more examples of network nonlocal correlations appeared in the literature ~\cite{Renou_2019,RB,Pozas,Sadra,Poderini2020,PRXQuantum.3.030342,PhysRevLett.130.190201,PhysRevLett.128.010403,ZHANG202122,wang2024experimental} (for a detailed review see~\cite{tavakoli2022bell}).  

The nonclassical behavior observed in these findings is validated under the assumption of fully independent sources. However, this assumption can be challenged. From a practical standpoint, it is conceivable that the various sources within a quantum network may exhibit some level of classical correlation; for example, these sources may require calibration and/or synchronization. The resilience of network nonclassicality to correlations among sources has been investigated in~\cite{Supic2020}. Therein the authors demonstrate that certain well-known instances of network nonlocality remain robust even in the presence of %minor 
correlations between the sources. Moreover, they introduce a case of quantum nonlocality without inputs within the triangle network that cannot be replicated with a network hidden variable model, unless the hidden variables are fully correlated.

In this work, we explore further theoretically and experimentally the example of quantum triangle nonlocality without inputs robust to arbitrary strong correlations among the sources. Initially, we highlight the prerequisite for observing this phenomenon, which necessitates the detection of perfectly correlated outputs between two pairs of parties, namely Alice and Charlie, and Bob and Charlie. Considering the challenges associated with achieving perfect correlations in experimental setups, we undertake a theoretical investigation into the behavior of these correlations under the influence of noise. Subsequently, we derive a Bell-like inequality that accommodates such imperfections. We experimentally demonstrate for the first time the violation of this derived inequality in a broad parameter space corresponding to non-fully-independent sources by exploiting the quantum state generated by a broadband integrated AlGaAs entangled photon source~\cite{Appas2022}. The presence of three distinct sources is emulated by demultiplexing the spectrum of the emitted quantum state into three pairs of frequency channels and by engineering it in an all-fibered architecture. The faithful simulation of a triangle quantum network architecture with the proposed system is assessed by demonstrating that the mutual information between states traversing different frequency channels approaches zero.  

The paper is structured as follows. In Section~\ref{sec:level2} we recapitulate the findings in~\cite{Supic2020} and present new results taking into account realistic, non-perfect correlations in two triangle links. In Section~\ref{exp} we describe the experimental implementation of the triangle quantum network architecture by using a broadband frequency multiplexed entangled photon pair source based on an AlGaAs Bragg reflection waveguide. We also demonstrate the violation of the Bell-like inequality relevant for this architecture. Finally in Section~\ref{conc} we offer conclusions and perspectives for future work.

%Exploring network nonlocality not only contributes to theoretical advancements but also holds practical implications for future quantum communication networks. 
\section{\label{sec:level2}Witnessing triangle inequality without inputs involving correlations among the sources}
\subsection{Triangle network with correlated sources}
In the configuration that interests us here, three parties, Alice, Bob, and Charlie, form a triangle network. Unlike the standard Bell scenario where inputs are necessary, here Alice, Bob, and Charlie produce outcomes, each consisting of two bits, labeled respectively $a = (a_B,a_C)$, $b= (b_A,b_C)$ and $c = (c_A,c_B)$, without receiving any inputs. The corresponding random variables are denoted by capital letters $(A_B,A_C,B_A...)$. By repeating the experiment many times the parties can collect correlation probabilities $p(a,b,c)$. We say that the experiment admits an explanation in terms of hidden variable models if the obtained correlation probabilities allow for the decomposition
\begin{equation}\label{Local_model_3sources}
p(a,b,c)=\sum_{\alpha,\beta,\gamma}p(\alpha,\beta,\gamma)p(a|\beta,\gamma)p(b|\alpha,\gamma)p(c|\alpha,\beta)
\end{equation}
where $\alpha$, $\beta$ and $\gamma$ are three hidden variables. If the sources comprising the triangle network are independent, the probability distribution $p(\alpha,\beta,\gamma)$ fully factorizes, \emph{i.e.} 
$p(\alpha,\beta,\gamma)=p(\alpha)p(\beta)p(\gamma)$. In the case of partial correlations among the sources, this factorization is no longer valid. The existing correlations can be taken into account by introducing a new hidden variable $\lambda$, which can influence the values of $\alpha$, $\beta$, and $\gamma$. We can thus rewrite the expression of $ p(a,b,c)$ in the case of partial dependence among the sources expressed by the following inequalities holding for all values of $\alpha$, $\beta$, $\gamma$ and $\lambda$: 
 % but not in an unrestricted way otherwise they could be entirely dependent on each other. It is essential to maintain some degree of independence that can be as small as possible as long as it is not equal to zero.
% This condition is expressed through these two constraints: 
\begin{align}\label{constraint1}
p(\alpha,\beta,\gamma|\lambda) &\geq \epsilon_1 p(\alpha)p(\beta)p(\gamma),  \\
p(\alpha,\beta,\gamma|\lambda) &\leq \epsilon_2(\alpha,\beta,\gamma,\lambda) p(\alpha)p(\beta)p(\gamma) ,\label{constraint2}
\end{align}   
where $\epsilon_1 \in (0,1]$ is a constant and $\epsilon_2(\alpha,\beta,\gamma,\lambda) \in [1,1/p(\alpha)p(\beta)p(\gamma)]$.

Given that the variable $\epsilon_1$ quantifies the degree of independence among the three sources, a local hidden variable model including the variable $\epsilon_1$ can then be defined. We will say that a distribution is $\epsilon_1$-trilocal if it admits a decomposition of the form 
\begin{equation}
p(a,b,c)=\sum_{\lambda,\alpha,\beta,\gamma}p(\lambda)p(\alpha,\beta,\gamma|\lambda)p(a|\beta,\gamma)p(b|\alpha,\gamma)p(c|\alpha,\beta)
\end{equation} 
where $p(\alpha,\beta,\gamma|\lambda)$ satisfies both bounds~\eqref{constraint1},\eqref{constraint2}.
If $\epsilon_1=1$ we recover the case of fully independent sources, thus obtaining the distribution defined in Eq.~(\ref{Local_model_3sources}). In~\cite{Supic2020}, the authors have theoretically demonstrated the existence of quantum correlations that cannot be explained by any local hidden variable model, even when the sources feature an arbitrary small level of independence, \emph{i.e.}, imposing only  $\epsilon_1>0$. This result was obtained by defining a Bell-like inequality for the triangle network and showing that it is possible to violate it. The first condition for validity of this Bell-like inequality is the observation of full correlations between outputs $a_C$ and $c_A$ on one side, and $b_C$ and $c_B$ on the other:
\begin{align}\label{perfcorr}
    p(A_C = C_A) = 1, \quad p(B_C = C_B) = 1. 
\end{align}

If the condition~\eqref{perfcorr} is satisfied, the inequality from~\cite{Supic2020} witnessing incompatibility with trilocal models has the following form
\begin{align} \nonumber
     \xi_1 &p(a_B = 0,b_A = 0,a_C = 0,b_C = 0)
     - \\ \nonumber- \xi_2 &\big(p(a_B= 0,b_A = 1,a_C = 0,b_C = 1)+ \\ \nonumber + &p(a_B = 1,b_A = 0,a_C = 1,b_C = 0)+\\ + &p(a_B = 0,b_A = 0,a_C = 1,b_C = 1)\big),
     \leq 0 \label{Bell_ineq_triangle}
\end{align}
where the parameters $\xi_1$ and $\xi_2$ are given by:
\begin{align}
    \xi_1&=\frac{\epsilon_1^3}{\epsilon^4_2}\min _{a_C:p(a_C)>0} p(a_C) \min_{b_C:p(b_C)>0} p(b_C)\\
       \xi_2&=\frac{\epsilon_2^3}{\epsilon^4_1}\max _{a_C:p(a_C)>0} p(a_C) \max_{b_C:p(b_C)>0} p(b_C),
\end{align}
and $p(a_C)$ and $p(b_C)$ are marginal probabilities of obtaining solely outputs $a_C$ and $b_C$, respectively. In what follows we present the quantum strategy violating inequality~\eqref{Bell_ineq_triangle}.

Charlie shares a maximally entangled state (a Bell pair) with both Alice, $\rho_{AC}$, and Bob, $\rho_{BC}$:
\begin{equation}\label{Perfect_State1}
       \rho_{AC}=\rho_{BC}=\proj{\phi^+} \hspace{0.5cm};\hspace{0.5cm}\ket{\phi^+}=\frac{1}{\sqrt{2}}(\ket{11}+\ket{00})
\end{equation}
and Alice and Bob share the state $\rho_{AB}=\ket{\psi_{\theta}}\bra{\psi_{\theta}}$, where 
\begin{equation}\label{Eberhard_state}
    \ket{\psi_{\theta}}=\frac{1}{\sqrt{1+\cos^2(\theta)}}[\sin\theta\ket{11}+\cos\theta(\ket{01}+\ket{10})]
\end{equation}

In this setup, Charlie measures both his photons in the computational basis $(0,1)$ and obtains the results $c_A$ and $c_B \in \{0,1\}$, respectively.
Similarly, Alice and Bob measure their photons correlated with Charlie in the same basis, obtaining the results $a_C$ and $b_C$, which exhibit perfect correlation with Charlie's outputs, \emph{i.e.}, $(a_C,b_C)=(c_A,c_B)$, thus satisfying condition~\eqref{perfcorr}.
Subsequently, the state shared between Alice and Bob, $\rho_{AB}$, is measured. The choice of the projection basis depends on the outcomes $(a_C,b_C)$. If $a_C=1$ ($b_C=1$), Alice (Bob) uses the computational basis $(0,1)$. Otherwise, if  $a_C=0$ ($b_C=0$), Alice (Bob) uses the basis defined as follows:
\begin{equation}\label{def_basisw0w1}
\begin{split}
    \ket{w_0}&=\sin\theta\ket{0}-\cos\theta\ket{1}\\
    \ket{w_1}&=\cos\theta\ket{0}+\sin\theta\ket{1},
\end{split}
\end{equation}
where $\theta$ is the angle introduced in Eq.~(\ref{Eberhard_state}).

The resulting quantum distribution $p_{\mathcal{Q}}^\theta(a_B,b_A,a_C,b_C)$ violates  inequality~\eqref{Bell_ineq_triangle}. Indeed, it can be easily verified that the conditional probability $p(a_B,b_A|a_C,b_C)$ satisfies the following conditions:
\begin{equation}\label{perfect_condition_quant_distribution}
    \begin{split}
        p(0,0|1,1)&=|\bra{00}\rho_{AB}\ket{00}|^2=0\\
        p(0,1|0,1)&=|\bra{w_01}\rho_{AB}\ket{w_01}|^2=0\\
        p(1,0|1,0)&=|\bra{1w_0}\rho_{AB}\ket{1w_0}|^2=0\\
        p(0,0|0,0)&=|\bra{w_0w_0}\rho_{AB}\ket{w_0w_0}|^2>0\\
    \end{split}  
\end{equation} 
leading to $I=\xi_1p(0,0,0,0) > 0$ for any $\epsilon_1 >0$, therefore violating inequality~(\ref{Bell_ineq_triangle}).
\begin{figure}
    %\centering
    \includegraphics[scale=0.08]{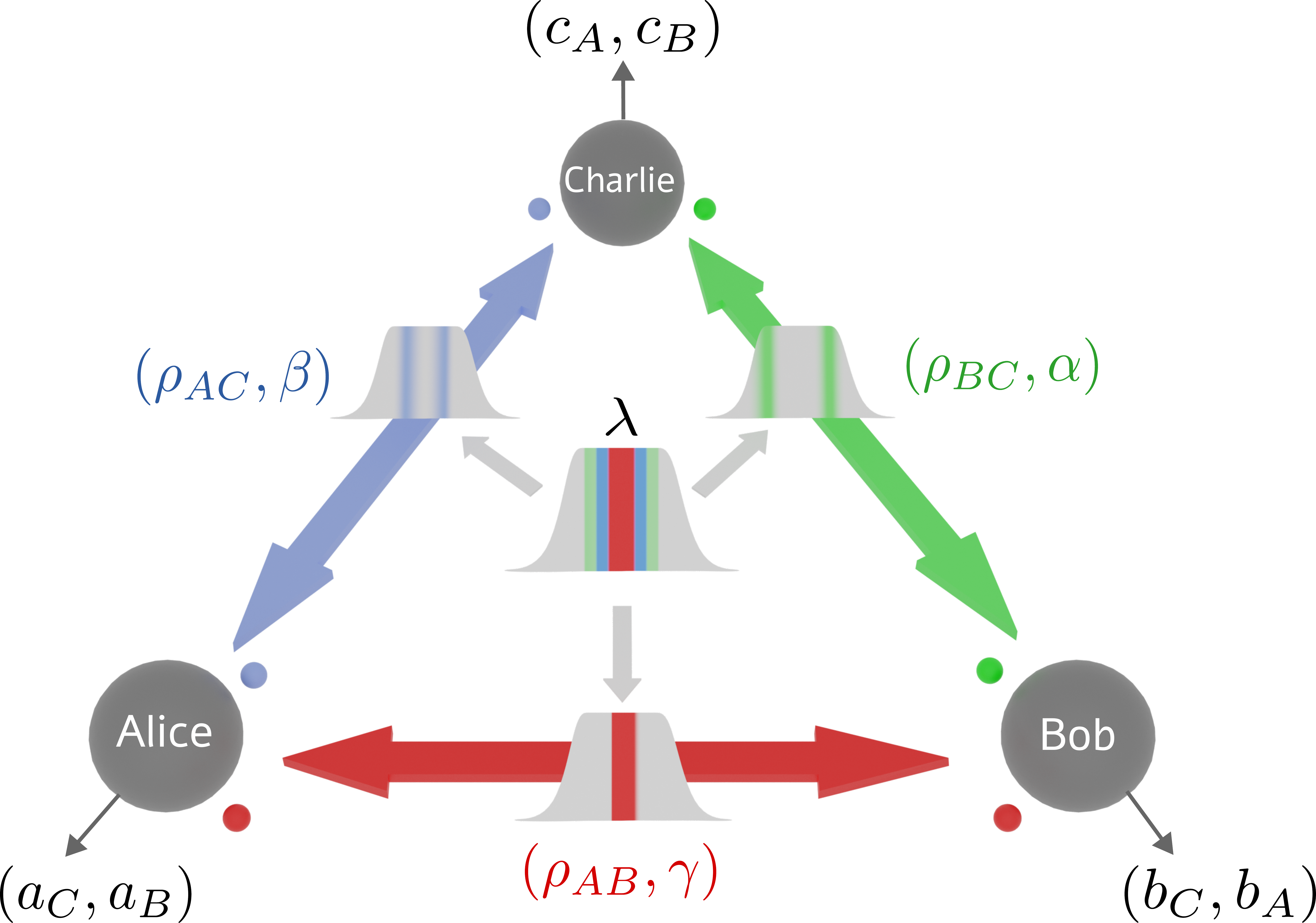}
    \caption{Scheme of a simulated triangle network using one physical source whose spectrum is multiplexed into three energy-matched frequency channels. The three states $\rho_{BC}$, $\rho_{AC}$, and $\rho_{AB}$ shared between Bob and Charlie, Alice and Charlie and Alice and Bob, respectively, are associated with a local hidden variable $\alpha$, $\beta$, and $\gamma$ respectively. The possible correlations between them are taken into account by introducing a fourth local hidden variable $\lambda$.}
    \label{fig:network}
\end{figure}

%\begin{equation}
 %   \ket{\psi}=\iint_{\omega_s\geq\omega_i} d\omega_sd\omega_i [C(\omega_s,\omega_i)\ket{HV}+C(\omega_i,%\omega_s)\ket{VH}]\ket{\omega_s\omega_i}
%\end{equation}

\subsection{A robust model for non-perfect sources} It is important to stress here that  perfect quantum correlations demanded in condition~\eqref{perfcorr} are unachievable under realistic conditions. In the following, we generalize inequality~(\ref{Bell_ineq_triangle}) to the case of imperfect sources. For this, we introduce the parameter $\Delta$, representing the probability that either Alice or Bob doesn't obtain the same output as Charlie:
\begin{equation}\label{Condition_imperfect_correlation}
    p(A_C=C_A, B_C=C_B)=1-\Delta
\end{equation} 
This leads to the following generalized inequality:
\begin{equation}\label{Bell_ineq_triangle_modif}
\begin{split}
     I= &\xi_1p(0,0,0,0)-\xi_2[p(0,1,0,1) \\
     &+p(1,0,1,0)+p(0,0,1,1)] \leq  \xi_1 \Delta  \\
\end{split}
\end{equation}
When the perfect correlation condition is satisfied, we have $I < 0$. However, when this condition is not met, $I < \mathcal{A}$, where $\mathcal{A}$ represents the algebraic bound--the maximum value achievable by $I$. This corresponds to the scenario where Alice and Bob always obtain $a_B=b_A=0$ while Charlie's outcomes, $(c_A,c_B)$, are different from $(0,0)$. This results in $p(0,0,0,0)=\Delta$ and $p(1,0,1,0)=p(0,1,0,1)=p(0,0,1,1)=0$, leading to $\mathcal{A}=\xi_1\Delta$. By defining the parameter $S_\Delta=I-\xi_1 \Delta$, we find that a $\epsilon_1$-trilocal distribution must satisfy, under the condition (\ref{Condition_imperfect_correlation}), the following inequality 
\begin{equation}\label{Inequality_S}
    S_\Delta \leq 0
\end{equation}
When $\Delta=0$, we retrieve the inequality (\ref{Bell_ineq_triangle})
A rigorous demonstration of this result is provided in Appendix~\ref{Inequality_Demo}.

\section{Experimental implementation of a triangle network using an AlGaAs source of broadband entangled photons}\label{exp}
\subsection{Experimental setup}
\begin{figure*}[t]
    %\centering
    \includegraphics[width=2\columnwidth]{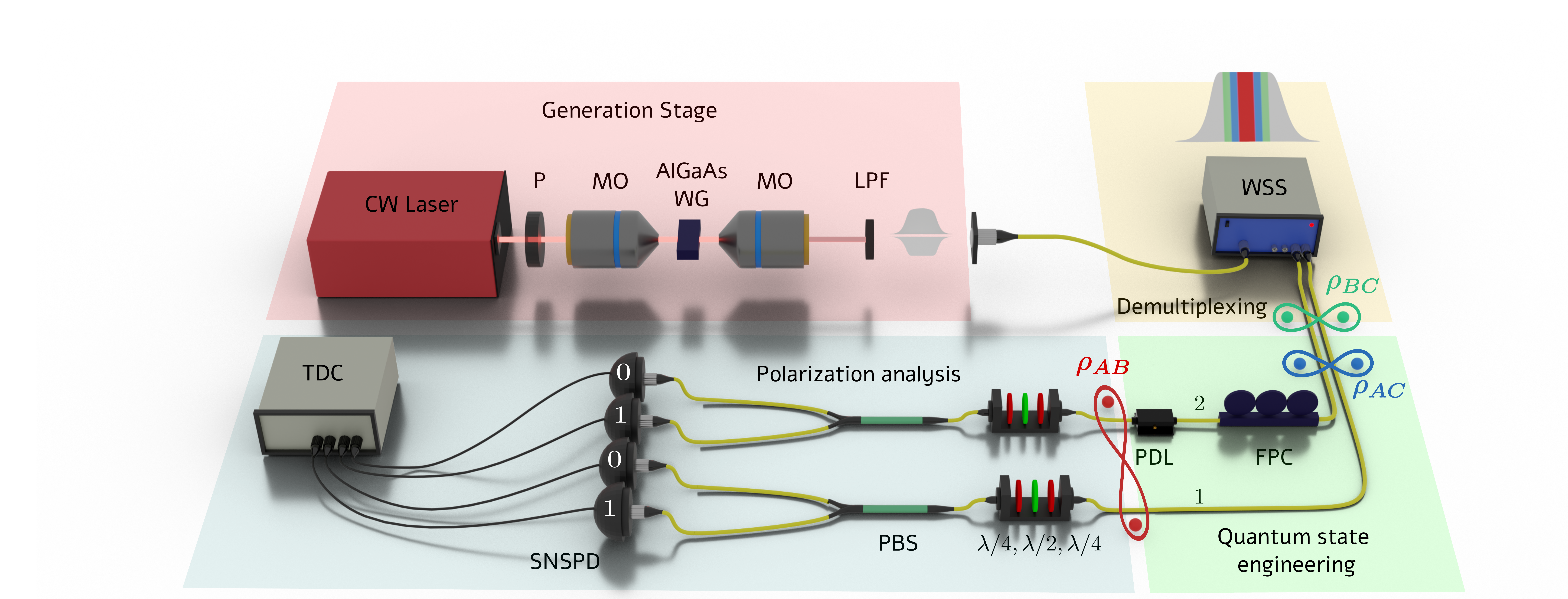}
    \caption{Sketch of the experimental setup for the implementation of the triangle network using a single source of broadband multiplexed entangled photon states, showing the generation, demultiplexing, quantum state engineering, and polarization analysis stages. The transmission window (in
    nm) for each filter is as follows:  red channel $[1544.8,1550.75] \cup [1538.5,1544.45]$, blue channel $[1550.9, 1553.9] \cup [1535.4, 1538.34]$, and green channel $[1554, 1557.1] \cup [1532.3, 1535.2]$. P: Polarizer; MO: Microscope objective; LPF: Long-Pass Filter; WSS: Waveshaper Selective Switch; FPC: Fiber Polarization Controller; PDL: Polarization Dependent Loss; PBS: Polarizing Beam Splitter; SNSPD: Superconducting Nanowire Single Photon Detector; TDC: Time-to-Digital Converter }
    \label{fig:setup}
\end{figure*}

We describe next the experimental setup used to simulate a triangle network architecture based on a single source generating broadband polarization-entangled photons. The spectrum of the generated quantum state is demultiplexed into three pairs of frequency-conjugate channels, as represented in Fig.~1. The experimental setup (see Fig.~\ref{fig:setup}) consists of four stages: i) photon pair generation, ii) demultiplexing, iii) quantum state engineering, and iv) polarization analysis. 

The source used in our setup consists of an AlGaAs Bragg-reflection waveguide operating at room temperature and generating pairs of signal and idler photons in the C telecom band through type-II spontaneous parametric down-conversion (SPDC)~\cite{Appas2022}. AlGaAs combines several assets in the landscape of material platforms for integrated quantum photonics~\cite{Baboux:23}: high second-order nonlinear effect~\cite{adachi1989optical}, direct bandgap compliant with electrical injection~\cite{PhysRevLett.112.183901}, and strong electro-optical effect leading to fast on-chip manipulation of quantum states~\cite{WANG201449}. Moreover, the very low bulk birefringence allows for a spectrally broadband emission of polarization entangled states circumventing the usual need to compensate for the group delay between orthogonally polarized photons~\cite{Kang:16}. This property is a valuable feature for the implementation of simple fibered architectures for quantum communications~\cite{Appas2021} as the one proposed in this work. 

In our experiment, a TE-polarized continuous-wave laser emitting at $\lambda_p=2\pi c /\omega_p= 772.3\nm $ is coupled, through a high numerical aperture microscope objective, to the AlGaAs waveguide generating orthogonally polarized signal and idler photons. In the following, the signal (idler) is associated with the highest (lowest) frequency $\omega_s$ ($\omega_i$) satisfying the condition of energy conservation: $\omega_i+\omega_s=\omega_p$.
Thus, the state generated by the source is described by:
\begin{equation}\label{State_Source}
\begin{split}
    \ket{\psi}=\iint_{\omega_i\leq \omega_s}d\omega_s d\omega_i [C(\omega_s,\omega_i)\ket{\omega_s,H}\ket{\omega_i,V}\\   
    +C(\omega_i,\omega_s)\ket{\omega_i,H}\ket{\omega_s,V}],
\end{split}
\end{equation} where $C(\omega_s,\omega_i)$ is the joint spectral amplitude, whose squared value gives the probability of producing a H(V)-polarized signal $\omega_s$ (idler $\omega_i$) photon. The generated photon pairs are collected with a second identical microscope objective, followed by a fiber collimator. A long-pass filter is used to eliminate the pump beam. 

The collected photon pairs are directed to a wavelength selective switch (WSS, model Finisar 4000s), where the signal $\omega_s$ and idler $\omega_i$ photons are filtered and separated into distinct output channels labeled 1 and 2, respectively. Three distinct filtering configurations are applied to provide the links between the three users; in Fig.~\ref{fig:setup} we have indicated in red, blue, and green the channels corresponding to the state shared between Alice and Bob $\rho_{AB}$, Alice and Charlie $\rho_{AC}$, and Bob and Charlie $\rho_{BC} $, respectively. Note that Alice always receives signal photons, Charlie receives always idler photons, whereas Bob receives the idler photon from $\rho_{AB}$ and the signal photon from $\rho_{BC}$. For Alice and Bob, we identify  $\ket{H}$ with  $\ket{0}$ and $\ket{V}$ with $\ket{1}$ , while for Charlie it is the contrary ($\ket{H}$ corresponds to $\ket{1}$ and $\ket{V}$ to $\ket{0}$).
\begin{figure*}[t]
    \centering
    \includegraphics[width=0.8\textwidth]{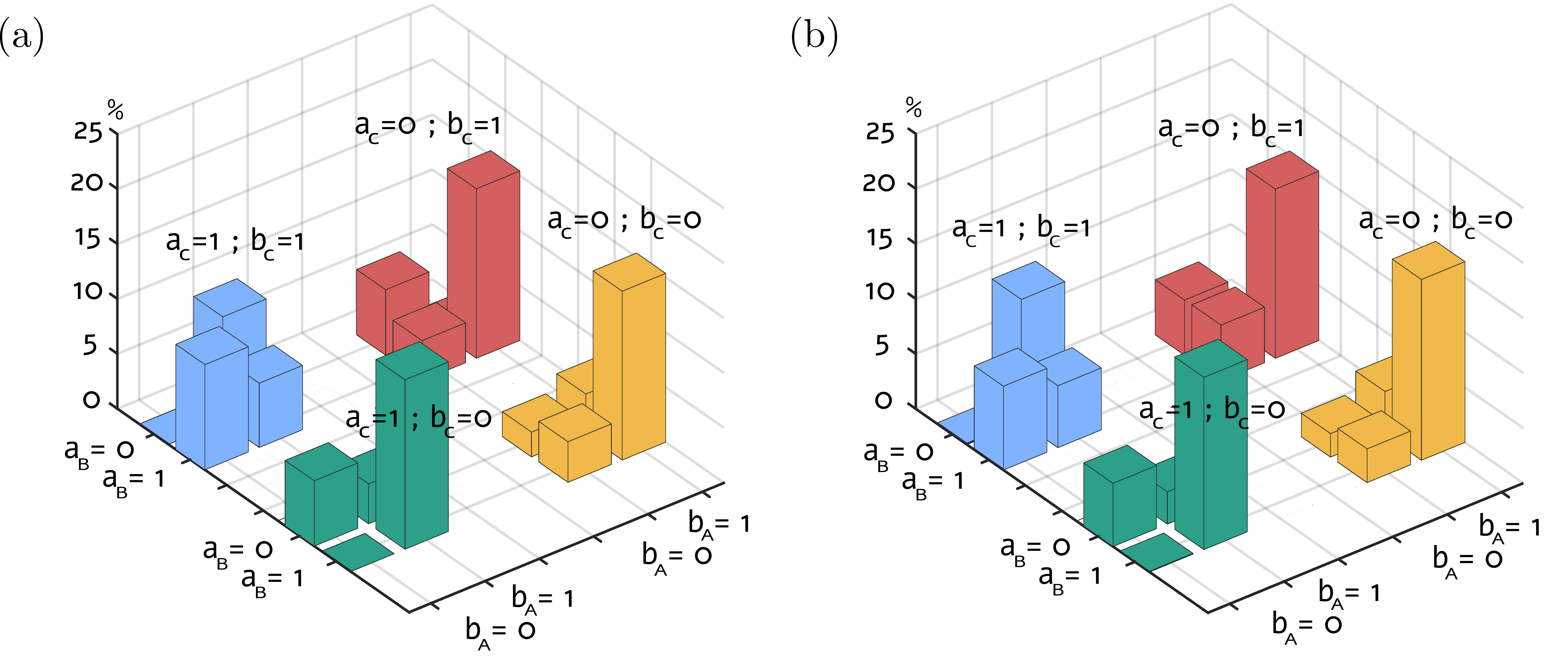}
-    \caption{Theoretical (a) and experimental (b) quantum distribution $p(a_B,b_A,a_C,b_C)$. The first is derived by considering perfect sources $\rho_{AB}$, $\rho_{AC}$, and $\rho_{BC}$ as defined in Eqs.\eqref{Perfect_State1} and \eqref{Eberhard_state}.  $(a_B,a_C)$ and $(b_A,b_C)$ represent the outcomes obtained by Alice and Bob, respectively. Each color corresponds to a basis projection for the state $\rho_{AB}$, defined by the output $(a_C,b_C)$.}
    \label{fig:Fritz}
\end{figure*}
In this case, at the WSS output, the state $\ket{\psi}$ can be written as:
\begin{equation}
\begin{split}
        \ket{\psi}=\int_{\Omega_1} d\omega_s \int_{\Omega_2}d\omega_i & [C(\omega_s,\omega_i)\ket{0}_1\ket{0}_2 \\
        +&C(\omega_i,\omega_s)\ket{1}_1\ket{1}_2],
\end{split}
\end{equation}
where $\Omega_1$ ($\Omega_2$) defines the frequency transmission of the filter applied to the signal (idler) photons. In this way, the states $\rho_{AC}$ and $\rho_{BC}$ are directly obtained at the output of the WSS.
The preparation of $\rho_{AB}$, instead, requires a manipulation of the quantum state directly produced by the source. For this, it is useful to show that the state $\ket{\psi_{\theta}}$ is equivalent to the partially entangled state:
\begin{equation}\label{partially_entangled_state}
    \ket{\psi_{r}}=\frac{1}{\sqrt{1+r^2}}(\ket{VH}+r\ket{HV})
\end{equation}
This can be demonstrated by defining as basis $\{\ket{0},\ket{1}\}$, the one rotated by an angle $\varphi_{A(B)}$ for Alice (Bob) with respect to the H and V axis (see Appendix~\ref{Proj_Basis}).
By doing so, we retrieve the expression of Eq.~(\ref{Eberhard_state}) with  $\theta$ given by:
\begin{equation}
        \sin{\theta}=\frac{r-1}{\sqrt{r^2-r+1} }\hspace{0.5cm} ; \hspace{0.5cm} \cos{\theta}=\sqrt{\frac{r}{r^2-r+1} }
\end{equation}

In our setup, we prepare the state $\ket{\psi_{r}}$  by introducing a polarization-dependent loss (PDL) emulator (OZ Optics model DTS0065) in channel 2. This passive optical component allows us to introduce (manually) adjustable losses depending on the polarization of the incoming light with respect to its eigenaxes.  The insertion of a fibered polarization controller before the PDL emulator permits to adjust $r^2$, that is the ratio between the transmission coefficients of the V and the H polarizations. In this way, the output state at the PDL can be written in the form of Eq.~\eqref{partially_entangled_state} (see the calculation in Appendix~\ref{PDL_calculation}). For $r=1$ (\emph{i.e.}, no losses introduced by the PDL emulator), we obtain the maximally entangled state $\ket{\psi}$ directly produced by the AlGaAs source; for $r \rightarrow 0$, the state is no more entangled; for $0<r<1$,  we obtain a partially entangled state.
The choice of the value of $\theta$ is done by maximizing the quantity $S_\Delta$,  and consequently the value of $p(0,0,0,0)$. Since, from equation \eqref{perfect_condition_quant_distribution}, we have $p(0,0,0,0)\propto p(0,0|0,0)=\frac{\cos^4\theta\sin^2\theta}{1+\cos^2(\theta)}$, its maximum is obtained for $\theta=38.2^\circ$, corresponding to an optimal value of the PDL $r_{\text{opt}}=0.464$. 

The signal and idler photons of the three different prepared states are then directed to a polarization analysis stage consisting of a fibered polarization rotator made of a set of free-space $\lambda/4$, $\lambda/2$, $\lambda/4$ waveplates, followed by a fibered polarizing beam splitter (PBS) and superconducting nanowire single-photon detectors (SNSPD) labeled either 0 or 1 (see Fig.~\ref{fig:setup}). Timestamps of the events are recorded with a time-to-digital converter (TDC).
%\begin{figure}[t]
%    \centering
    %\includegraphics[width=0.45\textwidth]{Figures/Tomography.pdf}
    %\caption{Reconstructed density matrices of the  three states $\rho_{AC}$, $\rho_{BC}$, and $\rho_{AB}$, shared by the users of the triangle network }
    %\label{fig:Tomg}
%\end{figure}
The density matrices of the three states shared in the network are reconstructed by quantum tomography (see Appendix~\ref{Quant_Tomog}). As expected, the density matrix corresponding to the partially entangled state exhibits unequal diagonal terms. A maximum fidelity $\mathcal{F}_{AB}=98\% \pm 0.2 \%$ is obtained for $r=0.4364$, in very good agreement with the target value $r_{\text{opt}}$. The fidelity with respect to the $\ket{\phi^+}$ Bell state for both $\rho_{AC}$ and $\rho_{BC}$ are also high: $\mathcal{F}_{AC}=97.7\% \pm 0.08 \%$, $ \mathcal{F}_{BC}=93.4\% \pm 0.2 \%$. The difference in fidelity between these two states can be attributed to birefringence and chromatic dispersion effects introduced by the source, which become non-negligible as one moves away from frequency degeneracy. However, this is not a problem in the context of our work; indeed, in their original proposal \cite{Supic2020} Supic \emph{et al.} suggest using a classically correlated state $\rho=\frac{1}{2}(\ket{00}\bra{00}+\ket{11}\bra{11})$ for the links that Charlie shares with Alice and Bob. Therefore, what is crucial is to achieve a strong correlation in the computational basis $(0,1)$, \emph{i.e.},  $(H,V)$; this is actually the case for both states since the measured correlation visibilities are $V_{HV}(\rho_{AC})=98.9\% \pm2\%$ and $V_{HV}(\rho_{BC})=98.8\% \pm2\%$.

\subsection{Bell-like inequality violation}
\begin{figure*}[t]
    \centering
    \includegraphics[width=2\columnwidth]{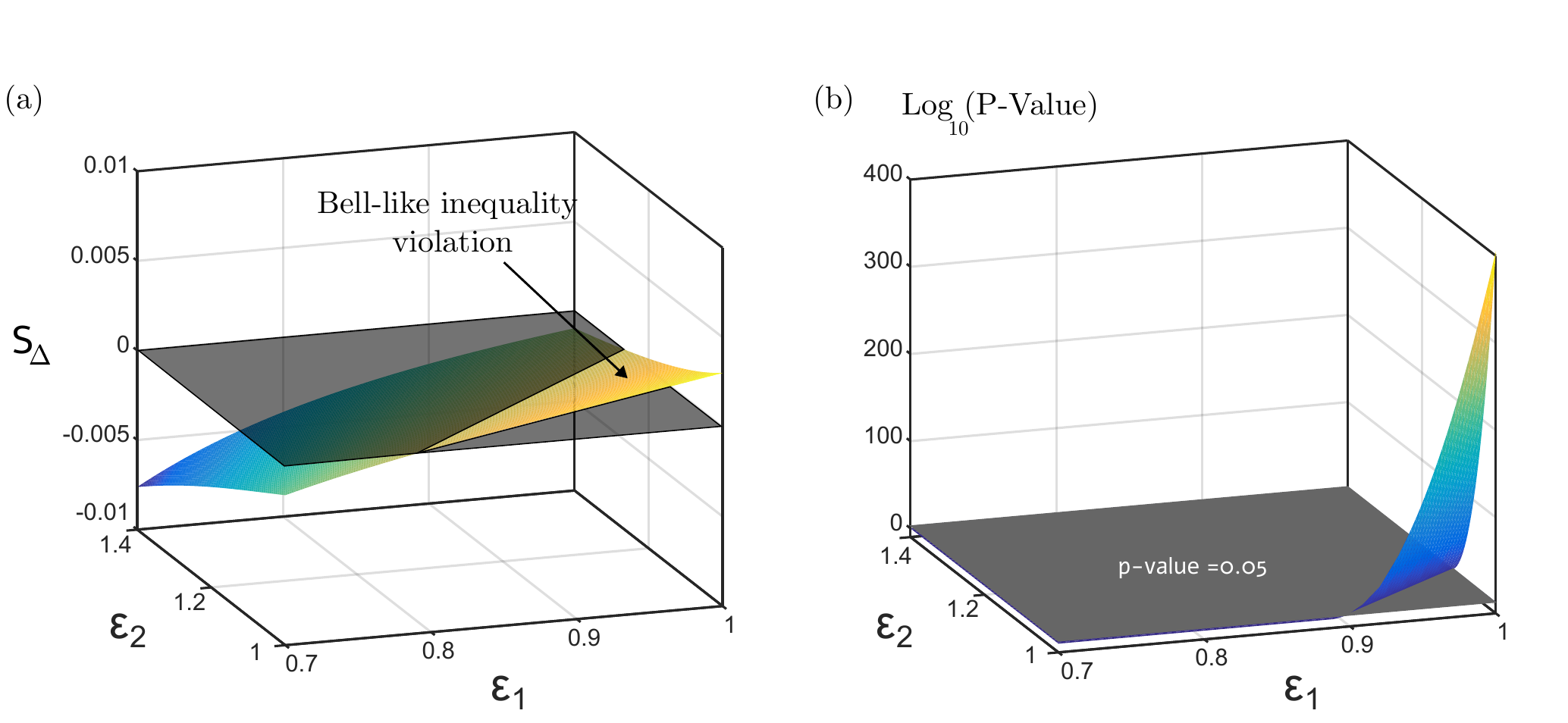}
    \caption{(a) $S_\Delta$ parameter deduced from the experimental results as a function of $\epsilon_1$ and $\epsilon_2$. The yellow surface above the grey plane $S_\Delta=0$ defines the region $\epsilon_1\times\epsilon_2$ where the Bell-like inequality is violated. (b) Corresponding statistical significance estimated via the p-value. A p-value lower than the threshold value of $0.05$ confirms the rejection of a LHV model. }
    \label{fig:Bell_Violation}
\end{figure*}
The reconstruction of the quantum probability distribution $p^\theta_{\mathcal{Q}}(a_B,b_A,a_C,b_C)$, is performed with three sequential measurements.

First, we select the state $\rho_{AC}$ %by applying the associated "blue" filter,
and share it between Alice and Charlie who measure their outcomes $(a_C,c_A)$ using the basis $(H,V)$. The measurement is done over an integration time of 8 hours leading to a total of 407~545 events in one of the four possible scenarios where detectors (0-0), (0-1), (1-0), or (1-1) click simultaneously.
 
%, resulting in 407 545 events. An event corresponds to two detector clicks, one for each partner, occurring within a time interval equal to the delay between the arrival of two photons from the same pair. These delays are estimated at $\delta t= \SI{486 \pm 324}{ps}$, $ \SI{14094 \pm 243}{ps}$, $ \SI{11259\pm 243}{ps} $, and $ \SI{3645 \pm 405}{ps}$, for scenarios where the detector pairs (0-0), (0-1), (1-0), (1-1) click, respectively.
 
Second, we repeat this procedure by selecting the state $\rho_{BC}$ and sharing it between Bob and Charlie. %after changing the filter from "blue" to "green" and substituting Alice with Bob in the setup.
An integration time of 8 hours results in 411~867 events $(b_C,c_B)$.

Third, we prepare the state $\rho_{AB}$ by selecting the corresponding frequency channels with the WSS and by introducing the PDL emulator in channel 2. As explained in Section~\ref{sec:level2}, four projections are performed: Alice and Bob either use the same basis or opposite basis. This results in four tables of events $(a_B,b_A)$: Table 1 for $(0,1) \otimes (0,1)$ , Table 2 for $(w_0,w_1) \otimes (w_0,w_1)$, Table 3 for $(0,1) \otimes (w_0,w_1)$, and Table 4 for $(w_0,w_1) \otimes (0,1)$, each one obtained over an integration time of 2 hours and containing respectively 108~286, 102~071, 101~857, 115~729 events. 
%From these four choices, we select the first output $(a_B,b_A)$ corresponding to the first output $(a_C,b_C)$.  For instance, we first obtained $(a_C=1,b_C=1)$, therefore we must pick the first output $(a_B,b_A)$ from the table corresponding to the projection $(0,1) \otimes (0,1)$. We proceed similarly for subsequent outputs, stopping until one of the six tables has been run completely. We end up with a table of length 399 410. Finally, the probability distribution $p^\theta_{\mathcal{Q}}(a_B,b_A,a_C,b_C)$ is estimated. Note that only hardware limitations oblige us to realize the measurements sequentially. Indeed, one can imagine using a 1x6 demux to create the six frequency channels of the networks and perform six-fold coincidences.\\
To obtain the complete data set $(a_B,b_A,a_C,b_C)$ allowing to test the Bell-like inequality, we must match the events $(a_B,b_A)$ with $(a_C,b_C)$ according to the protocol explained in Section~\ref{sec:level2}. For example, if $(a_C,b_C)=(1,1)$, then $(a_B,b_A)$ is selected from the first table. Similarly, $(a_C,b_C)=(0,0)$, $(1,0)$, or $(0,1)$ corresponds to selecting $(a_B,b_A)$ from the second, third, or fourth table, respectively. We end up with a final table containing 399~410 events.
Note that the availability of a 1x6 demultiplexer and six (instead of four) single-photon detectors would allow us to measure two-fold coincidences between the six pairs of channels simultaneously thus avoiding the sequential data collection.

In Fig.~\ref{fig:Fritz}, we show our experimental results (right panel) together with the corresponding theoretical results (left panel) obtained using the ideal target states $\ket{\phi^+}$ and $\ket{\psi_{\theta}}$.
Overall, the experimental results are in very good agreement with theoretical predictions. 
For instance, we obtain values very close to zero for the probabilities $p(0,1,0,1)=6.785\times 10^{-5}$, $p(1,0,1,0)=6.785\times 10^{-5}$, and $p(0,0,1,1)=1.4 \times 10^{-3}$, as expected from relations~(\ref{perfect_condition_quant_distribution}). The difference observed between the experimental and theoretical values for some projections, such as $p(0,1,1,1)$ and $p(1,0,1,1)$ is attributed to a slight misalignement of the axis of the PDL emulator. A detailed analysis in Appendix~\ref{Alignment_PDL} shows that this misalignment, limited to the range $[-3^\circ,4^\circ]$, does not affect the value of $S_\Delta$ and the validity of our results.

%. Importantly, it is shown that the probabilities taken into account for the calculation of $S_\Delta$ do not appear to be sensitive to an angle deviation within this range, meaning that even if we were perfectly aligned, it would not affect the value of $S_\Delta$.

We then calculate the coefficient $\Delta$ corresponding to the probability that either Alice or Bob doesn't get the same output as Charlie (see Eq.~(\ref{Condition_imperfect_correlation})). As expected, we obtain a small value ($0.62\%$), confirming the high level of correlation of the measurement outcomes for the two states $\rho_{AC}$ and $\rho_{BC}$. In Fig.~\ref{fig:Bell_Violation}.(a), we show the value of $S_\Delta$ calculated from our measurements as a function of the two free variables $\epsilon_1$ and $\epsilon_2$: the region above the plane $z=0$ corresponds to the zone for which inequality~(\ref{Inequality_S}) is violated, clearly showing that this is possible for non-independent sources (\emph{i.e.}, sources for which $\epsilon_1<1$). %, delimited as $\epsilon_1 \in [0.79 \, , \,1]$ and $\epsilon_2 \in [1 \, , \, 1.26]$.demonstrating the existence of quantum nonlocality even in the case of non-independent sources (e.g. $\epsilon_1<1$). 

We point out that the values of $\epsilon_1$ and $\epsilon_2$ are not observable quantities, but they are used to describe a network hidden variable model we aim to refute. One way to get an indication of their value is through their relation to the mutual information existing between the random variables $A_B$, $A_C$, $B_A$, $B_C$, $C_A$, and $C_B$ (see Appendix~\ref{Mut_Inf} for details). As the measured value of the mutual information between non-correlated variables is of the order of $10^{-7}$, by assuming that equations~\eqref{constraint1} and~\eqref{constraint2} hold uniformly for every $\lambda$, we can infer that both $\epsilon_2-1$ and $1-\epsilon_1$  are of the order of $10^{-7}$. We thus conclude that for such a network hidden variable model the results shown in Fig.\ref{fig:Bell_Violation}.a) witness nonlocality.

Finally, as finite statistics are relevant in our experiment, it is important to quantify the statistical level of confidence with respect to a local hidden variable (LHV) model. For this, as proposed in~\cite{Elkouss2016}, we use as statistical tool the p-value, representing the probability that the observed data can be explained by a LHV model (see Fig.~\ref{fig:Bell_Violation}.(b). In our case, the p-value depends on the value of $\epsilon_1$ and $\epsilon_2$ (see Appendix~\ref{P_value}). 
When $\epsilon_1$ and $\epsilon_2 \rightarrow 1$, the p-value $\rightarrow 10^{-400}$, thus rejecting the hypothesis of a LHV model. Fig.~\ref{fig:Bell_Violation}.(b) also shows the p-threshold value $0.05$, commonly used to quantify the statistical significance of the results, confirming the observation of quantum nonlocality in our triangle network. 
%The new region of violation, with strong evidence against local models, is reduced to $\epsilon_1 \in [0.89 \, , \,1]$ and $\epsilon_2 \in [1 \, , \, 1.12]$. 

\section{Conclusion}\label{conc}
In this work, we explored both theoretically and experimentally quantum nonlocality without inputs in triangle networks robust to arbitrarily strong correlations among the sources. We theoretically analyzed how these correlations are affected by noise and derived a Bell-like inequality accounting for these imperfections. We then experimentally showcased for the first time the violation of this inequality in a parameter range that is compatible with non-fully-independent sources using energy-matched channels of an AlGaAs entangled photon source. We stress that all the stages of the experiment including demultiplexing, state engineering, polarization stage analysis, and detection are based on off-the-shelf fibered components resulting in a stable and easily reconfigurable system fully compatible with current fiber networks and well-suited for future applications outside the laboratory. 

A possible future upgrade of our architecture includes pigtailing the AlGaAs chip to obtain a fully fibered plug-and-play setup. Further progress is possible in several ways: the full control over channel frequency, width, and transmission offered by the WSS opens the possibility of tuning the degree of correlation among the sources via a controllable overlap of the distributed frequency channels so as to investigate the impact of source correlations on the network nonlocality. In addition, the broadband polarization entanglement of the state generated by the AlGaAs chip enables the investigation of more complex network topologies involving more than three parties by using a larger number of frequency channels. For example, it has been shown that the square network can also exhibit quantum nonlocality with partial correlations between the sources (see Supplemental Material of~\cite{Supic2020}). 

Such experimental investigations require adapting the underlying theoretical framework by identifying suitable Bell-like inequalities and taking into account relevant practical imperfections. From a fundamental point of view, removing implicit assumptions and experimental loopholes and hence leading to device-independent implementations is a significant but important challenge, opening the way to certification of the quantum network resources used in the corresponding applications.

\section{Acknowledgments}\label{Ack}
This work was supported by Defence Innovation Agency under grant ANR-19-ASTR- 0018 01), the French
RENATECH network, the Plan France 2030 through the project ANR-22-PETQ-0011, the Paris Ile-de-France Region in the framework of DIM SIRTEQ through the Project STARSHIP, and the European Union’s Horizon Europe research and innovation program under the project ”Quantum Security Networks Partnership” (QSNP, grant agreement No 101114043). O.M. acknowledges Labex SEAM (Science and Engineering for Advanced Materials and devices), ANR-10-LABX-0096 and ANR-18-IDEX-0001 for financial support.

\appendix
\section{Bell-like inequality derivation}\label{Inequality_Demo}
To construct the noise-robust Bell-like inequality, we start from a general expression of $I$:
\begin{equation}\label{Definition_I}
	I=\sum_{a,b}\left(\xi_1w^+_{a,b}-\xi_2w^-_{a,b}\right) p(a_B,b_A,a_C,b_C)
\end{equation}
where $w^+_{a,b}$ and $w^-_{a,b}$ are real non-negative coefficients that define a standard Bell inequality:
\begin{equation}
	\sum_{a,b}\left(w^+_{a,b}-w^-_{a,b}\right) p(a_B,b_A|a_C,b_C)\leq L
\end{equation} with $L$ being a local bound.
The probability distribution $p(a_B,b_A,a_C,b_C)$ can be written as follows:
\begin{align}
		p(a_B,b_A,a_C,b_C)&=\sum_{c_Ac_B}p(a_B,b_A,a_C,b_C,c_A,c_B)\\
		&=p_1(a_B,b_A,a_C,b_C)+p_2(a_B,b_A,a_C,b_C) \label{split_prob}
\end{align}
where $p_1(a_B,b_A,a_C,b_C)$ corresponds to the case where Charlie obtains the same output as Alice and Bob, i.e. $(c_A,c_B)=(a_C,b_C)$, and is equal to
\begin{equation}
	p_1(a_B,b_A,a_C,b_C)=p(a_B,b_A,a_C,b_C,c_A=a_C,c_B=b_C)
\end{equation}
The second term, $p_2(a_B,b_A,a_C,b_C)$, corresponds to the opposite cases, i.e $(c_A,c_B)\not=(a_C,b_C)$, and is equal to
\begin{equation}
	p_2(a_B,b_A,a_C,b_C)=\sum_{(c_A,c_B)\not=(a_C,b_C)}p(a_B,b_A,a_C,b_C,c_A,c_B)
\end{equation}
By substituting the expression \eqref{split_prob} into \eqref{Definition_I}, we obtain
\begin{equation}
		I=I_1+I_2
\end{equation}
where
\begin{equation}
		I_1=\sum_{a,b}\left(\xi_1w^+_{a,b}-\xi_2w^-_{a,b}\right) p_1(a_B,b_A,a_C,b_C)
\end{equation}
and 
\begin{equation}
	I_2=\sum_{a,b}\left(\xi_1w^+_{a,b}-\xi_2w^-_{a,b}\right) p_2(a_B,b_A,a_C,b_C)
\end{equation}
In \cite{Supic2020}, it is demonstrated that, under condition \eqref{perfcorr}, $I$ satisfies the following inequality :
\begin{equation}
	I \leq \xi_1\xi_2 L
\end{equation}
Thus, by definition, $I_1$ also satisfies it	 
\begin{equation}
	I_1 \leq \xi_1\xi_2 L
\end{equation}
Concerning $I_2$, since the condition of perfect correlations is no longer valid, there is no specific bound. Consequently, the maximum value achievable by $I_2$ is given by the algebraic bound $\mathcal{A}_2$
\begin{equation}
	I_2 \leq  \mathcal{A}_2
\end{equation}
Any distribution, whether local or quantum, must satisfy this inequality.
This leads to the following generalized inequality:
\begin{equation}\label{General_Ineq_Triangle_Network2}
		I \leq \xi_1\xi_2 L +\mathcal{A}_2
\end{equation}
In our case, $I$ has the form of \eqref{Bell_ineq_triangle}. Consequently, 
\begin{equation}
    \begin{aligned}
	I_{1(2)}= &\xi_1p_{1(2)}(0,0,0,0)-\xi_2[p_{1(2)}(0,1,0,1)+\\  
    & p_{1(2)}(1,0,1,0)+p_{1(2)}(0,0,1,1)] 
    \end{aligned}
\end{equation}
Reaching the algebraic bound $\mathcal{A}_2$ corresponds to the scenario where Alice and Bob always obtains $a_B=a_C=b_A=b_C=0$ and Charlie $(c_A,c_B)\not=(0,0)$, resulting in
\begin{equation}
	\mathcal{A}_2=\xi_1 p_2(0,0,0,0)
\end{equation}
Since the probability of having $(C_A,C_B)\not=(A_C,B_C)$ is equal to $\Delta$, then we obtain
\begin{equation}
	\mathcal{A}_2  = \xi_1 \Delta
\end{equation}
Knowing that in our case $L=0$, the inequality \eqref{General_Ineq_Triangle_Network2} becomes
\begin{equation}
		I\leq  \xi_1 \Delta
\end{equation}
which concludes the demonstration of \eqref{Inequality_S}.
\section{Polarization Basis Projection}\label{Proj_Basis}
The basis $\{\ket{0},\ket{1}\}$ used by Alice (Bob) when measuring the state $\rho_{AB}$ is defined as follows:
\begin{equation}\label{def_basis_01}
\begin{split}
    \ket{0}^{A(B)}&=\sin\varphi_{A(B)}\ket{H}+\cos\varphi_{A(B)}\ket{V}\\
    \ket{1}^{A(B)}&=\cos\varphi_{A(B)}\ket{H}-\sin\varphi_{A(B)}\ket{V}
\end{split}
\end{equation}
By choosing angles $\varphi_{A}$ and $\varphi_{B}$ that satisfy the following relations:
\begin{equation}\label{relation_phi_alpha}
    \sin{\varphi_B}=\frac{1}{\sqrt{1+r}} \hspace{0.22cm} ; \hspace{0.22cm} \cos{\varphi_B}=\frac{\sqrt{r}}{\sqrt{1+r}} \hspace{0.22cm} ; \hspace{0.22cm} \varphi_A = \varphi_B-\frac{\pi}{2}
\end{equation}
 the state $\ket{\psi_r}$ of Eq.~\eqref{partially_entangled_state} becomes equivalent to the state $\ket{\psi_\theta}$ of Eq.~\eqref{Eberhard_state}.
\begin{figure}[h]
    \centering
    \includegraphics[width=0.25\textwidth]{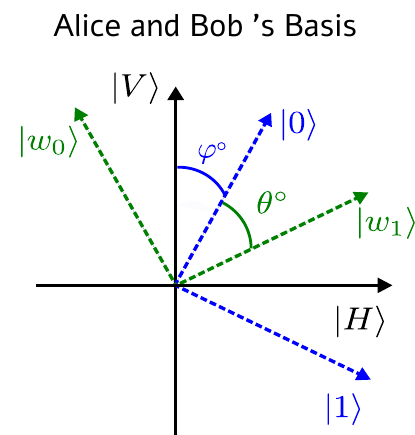}
    \caption{Polarization basis used by Alice and Bob for measuring the state $\rho_{AB}$. The angle $\varphi$ corresponds to $\varphi_{A(B)}$ for Alice (Bob).}
    \label{fig:basis}
\end{figure}

\section{Effect of PDL on the generated quantum state}\label{PDL_calculation}
%\begin{equation}
%	\ket{\psi}_{PDL}=\frac{1}{N}\int_{\Omega_1}d\omega_s \int_{\Omega_2} d\omega_i [\sqrt{t_H}C(\omega_s,\omega_i)\ket{HV}_{12}+\sqrt{t_V}C(\omega_i,\omega_s)\ket{VH}_{12}] 
%\end{equation}
The corresponding density matrix of the state at the WSS output is $\Tilde{\rho}=\ket{\psi}\bra{\psi}$, where
\begin{equation}
\begin{split}
        \ket{\psi}=\int_{\Omega_1} d\omega_s \int_{\Omega_2}d\omega_i [&C(\omega_s,\omega_i)\ket{H}_1\ket{V}_2 \\
        +&C(\omega_i,\omega_s)\ket{V}_1\ket{H}_2]
\end{split}
\end{equation} 
After tracing out the frequency, the reduced density matrix is expressed as:
\begin{equation}
\begin{split}
    \rho&=\frac{1}{N}\iint d\omega_1d\omega_2 \bra{\omega_1}\bra{\omega_2}\Tilde{\rho}\ket{\omega_1}\ket{\omega_2}   \\ 
    &=\alpha\ket{HV}\bra{HV}+D\ket{HV}\bra{VH}\\
    &+D^*\ket{VH}\bra{HV}+\beta\ket{VH}\bra{VH}
    \end{split}
\end{equation}
where 
\begin{align}
    \alpha&=\frac{1}{N}\int_{\Omega_1} d\omega_1 \int_{\Omega_2} d\omega_2|C(\omega_1,\omega_2)|^2\\
    \mathcal{D}&=\frac{1}{N}\int_{\Omega_1}d\omega_1 \int_{\Omega_2}d\omega_2C(\omega_1,\omega_2)C^*(\omega_2,\omega_1)\\
    \beta&=\frac{1}{N}\int_{\Omega_1} d\omega_1 \int_{\Omega_2}d\omega_2 |C(\omega_2,\omega_1)|^2
\end{align}
with $N=\int_{\Omega_1} d\omega_1 \int_{\Omega_2} d\omega_2 (|C(\omega_1,\omega_2)|^2+|C(\omega_2,\omega_1)|^2)$.\\
The action of the PDL emulator on the photon pair is given by :
\begin{align}
    \ket{HV} &\longrightarrow \sqrt{t_V}\ket{HV}+\sqrt{1-{t_V}}\ket{H0}\\
    \ket{VH} &\longrightarrow \sqrt{t_H}\ket{VH}+\sqrt{1-{t_H}}\ket{V0}
\end{align}
The photon $H$ ($V$) is transmitted with a probability of $t_H$ ($t_V$) and lost with a probability of $1-t_H$ ($1-t_V$).
Since we are interested in detecting coincidences, the vacuum term $\ket{0}$ can be neglected; the reduced density matrix becomes: 
\begin{equation}
    \begin{split}
    \rho_c=\frac{1}{N'}(&t_V \alpha \ket{HV}\bra{HV}+\sqrt{t_Ht_V}D\ket{HV}\bra{VH}\\
    +&\sqrt{t_Ht_V}D^*\ket{VH}\bra{HV}+t_H\beta \ket{VH}\bra{VH})
    \end{split}
\end{equation}
where $N'=t_V\alpha+t_H\beta$.\\
If we assume a symmetric joint spectral amplitude, i.e. $C(\omega_1,\omega_2)=C(\omega_2,\omega_1)$, we get $\alpha=\beta=D=1/2$
and thus:
\begin{equation}
    \begin{split}
    \rho_c=\frac{1}{1+r^2}(&\ket{VH}\bra{VH}+r\ket{HV}\bra{VH}\\
    +&r\ket{VH}\bra{HV}+r^2\ket{HV}\bra{HV})
    \end{split}
\end{equation}\\
with $r=\sqrt{t_V/t_H}$. Finally, we get a partially entangled state $\rho_c=\ket{\psi_r}\bra{\psi_r}$ where $\ket{\psi_r}$ is defined in Eq.~(\ref{partially_entangled_state}).\\

\section{Quantum Tomography}\label{Quant_Tomog}
\begin{figure}[h]
    \centering
    \includegraphics[width=0.45\textwidth]{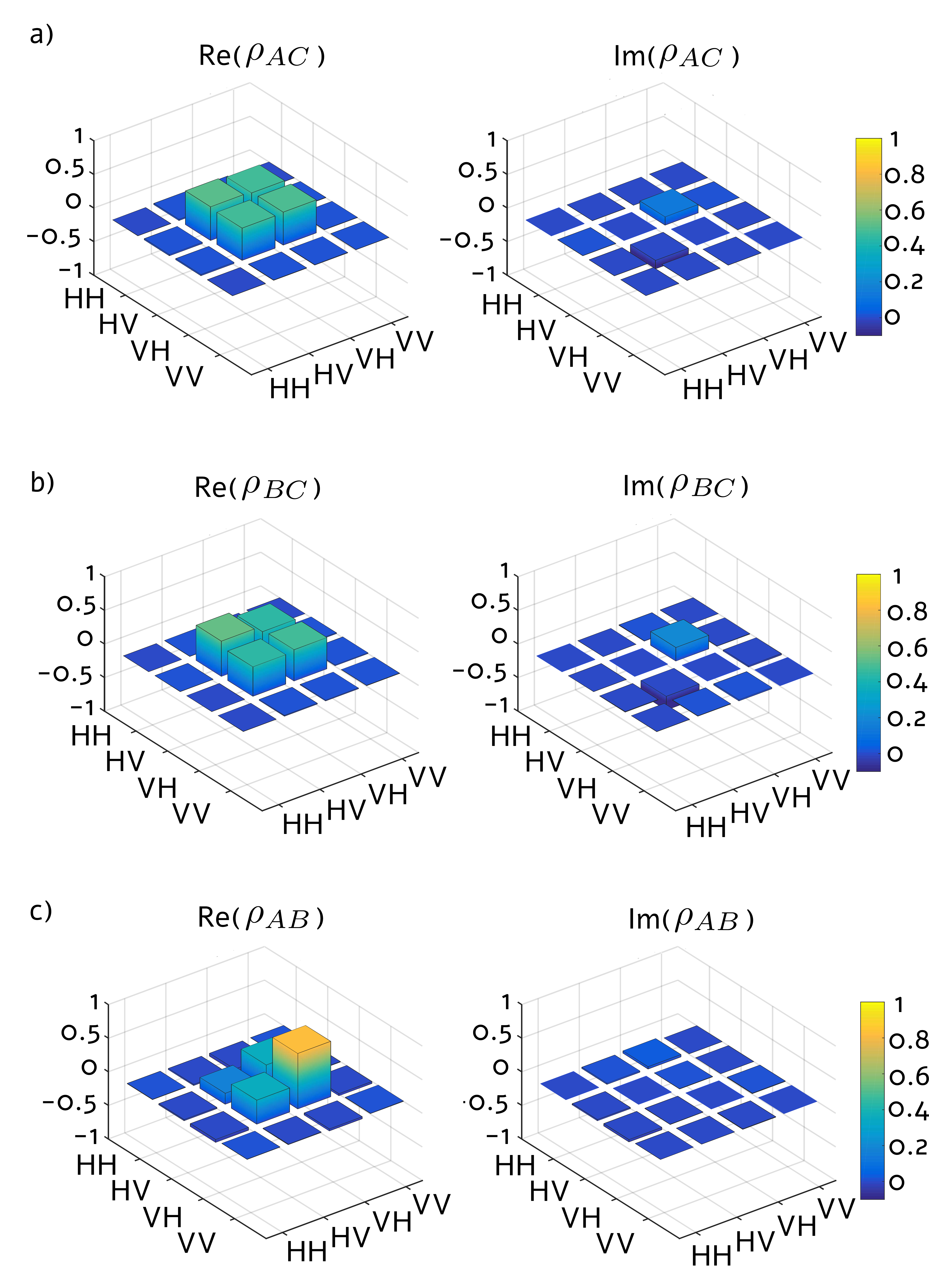}
    \caption{Reconstructed density matrices of the  three states $\rho_{AC}$, $\rho_{BC}$, and $\rho_{AB}$, shared by the users of the triangle network.}
    \label{fig:Tomg}
\end{figure}

The reconstructed density matrices of the three states $\rho_{AC}$, $\rho_{BC}$, and $\rho_{AB}$ are shown in Fig.~\ref{fig:Tomg}. The method used is based on maximum likelihood estimation, as proposed in \cite{PhysRevA.64.052312}.

\section{Mutual Information}\label{Mut_Inf}
The mutual information of two random variables $X$ and $Y$ is defined as :
\begin{equation}
I(X,Y)=\sum_{x,y}P(x,y) \log\frac{P(x,y)}{P(x)P(y)}
\end{equation}
When $X$ and $Y$ are independent of each other, the mutual information is zero since $P(x,y)=P(x)P(y)$. \\
In our network, six random variables are involved: $A_C$, $A_B$, $B_C$, $B_A$, $C_A$, and $C_B$. The different values of the mutual information are given in Table~\ref{Tab:Mutual_inf}. Since $A_C$ and $B_C$ are almost perfectly correlated with $C_A$ and $C_B$ respectively, $I(A_C,C_A)$ and $I(B_C,C_B)$ should then be close to $\log(2)\approx 0.3$. The fact that $I(A_C,B_C)$ and $I(C_A,C_B)$ are very small ($\approx 10^{-7}$) proves that the quantum states shared by Alice and Bob with Charlie are independent. Since the choice of basis for measuring the state $\rho_{AB}$ depends on the values of $A_C$ and $B_C$, the random variables $A_B$ and $B_A$ are supposed to be slightly correlated with the latter, which explains why $I(A_B,A_C)$ and $I(B_A,B_C)$ are of the order of $10^{-3}$.

\begin{table}[ht]
    \centering
    \begin{tabular}{|c|c|c|c|c|c|}
    \hline
         &  $A_C$ & $B_A$ & $B_C$ & $C_A$ & $C_B$ \\ \hline
         $A_B$  & $0.0092$ & $0.0079$  & $2.5 \times 10^{-4}$ & $0.0091$ & $2.6 \times 10^{-4}$ \\ \hline
         $A_C$  & - & $1.2 \times 10^{-4}$ & $7.8 \times 10^{-7}$ & $0.291$ & $7.68 \times 10^{-7}$ \\ \hline
         $B_A$  & - & - & $0.0049$ & $1.19 \times 10^{-4}$ & $0.0049$ \\ \hline
         $B_C$  & - & - & - & $7.8 \times 10^{-7}$ & $0.2927$ \\ \hline
         $C_A$  & - & - & - & - & $7.67 \times 10^{-7}$ \\ \hline
    \end{tabular}
    \caption{Two-by-two mutual information values calculated between the six variables involved in the network: $A_C$, $A_B$, $B_C$, $B_A$, $C_A$, and $C_B$.}
    \label{Tab:Mutual_inf}
\end{table}

\section{P-Value}\label{P_value}
The p-value tool quantifies the probability that the statistic derived from experimental data can be reproduced if the null hypothesis, represented by the local model in our case, is true.
Let us define a random variable $(E_i)_{1 \leq i\leq n}$ as $E_i=(a_B,b_A,a_C,b_C)$ representing the recorded events. When calculating the parameter $S$, only a dataset, which can be divided into three sets $\mathcal{A}_0$, $\mathcal{A}_1$, and $\mathcal{A}_2$, is considered with
 \begin{equation}
 \begin{split}
    \mathcal{A}_0&=\{(a_B,b_A,a_C,b_C)| (a_C,b_C)\neq (c_A,c_B) \}\\
    \mathcal{A}_1&=\{(0,0,1,1),(0,1,0,1),(1,0,1,0)\}\\
    \mathcal{A}_2&=\{(0,0,0,0)\}
 \end{split}
\end{equation}
The inequality test can be seen as a win/lose game. Let $(X_i)_{1\leq i\leq n}$ be a random variable defined as follows:
\begin{equation}
    X_i = \left\{
        \begin{array}{ll}
            1 & \text{if} \hspace{0.1cm} E_i \in \mathcal{A}_2  \hspace{0.1cm}  \text{and} \hspace{0.1cm} E_i \notin \mathcal{A}_0 \\
            0 & \text{if} \hspace{0.1cm}E_i \in \mathcal{A}_1 \cup \mathcal{A}_0
        \end{array}
    \right.
\end{equation}

The winning condition is reached when $X_i=1$. 
The optimal winning probability that can be achieved using an LHVM is denoted by $\beta_{win}$. After $n$ trials, the probability of winning at least $c$ times, equal to the p-value, is given, as demonstrated in \cite{Elkouss2016}, by:
\begin{equation}\label{p-value}
    \text{p-value}= \sum_{i=c}^{n} \binom{n}{c} \beta_{win}^i (1-\beta_{win})^{n-i}
\end{equation}
To derive the expression of $\beta_{win}$, we start by writing the inequality in this form :
\begin{equation}
    \xi_1 p(X_i=1)-\xi_2 p(X_i=0) \leq 0
\end{equation}
Knowing that $ p(X_i=1)+ p(X_i=0) = 1 $, we get an upper bound :
\begin{equation}\label{condition_proba_LHVM}
    p(X_i=1) \leq \frac{\xi_2}{\xi_2+\xi_1}
\end{equation} 
A probability distribution that an LHV model can reproduce must verify the constraint~(\ref{condition_proba_LHVM}). Therefore, $\beta_{win}=\xi_2/(\xi_2+\xi_1)$.\\
In our case, only $n=12045$ events out of a total number of $399 410$  contribute to the calculation of $S$, with $c=8481$ corresponding to the number of times we get $X_i=1$. 
Inserting the values of $c$, $n$, and $\beta_{win}$ in Eq.~(\ref{p-value}), leads to the p-value; for the sake of simplicity, we finally used an analytical expression to calculate numerically the p-value
\cite{Storz2023}:
\begin{equation}
    \text{p-value} \leq \frac{1}{\sqrt{2\pi}}\frac{\sqrt{n}}{\sqrt{c(n-c)}}\frac{(\frac{n \beta_{win}}{c})^c(\frac{n(1-\beta_{win})}{(n-c)})^{n-c}}{1-\frac{\beta_{win}(n-c)}{(1-\beta_{win})(c+1)}}
\end{equation}

\section{PDL Axis Alignment}\label{Alignment_PDL}
\begin{figure}[t]
    \centering
    \includegraphics[width=0.42\textwidth]{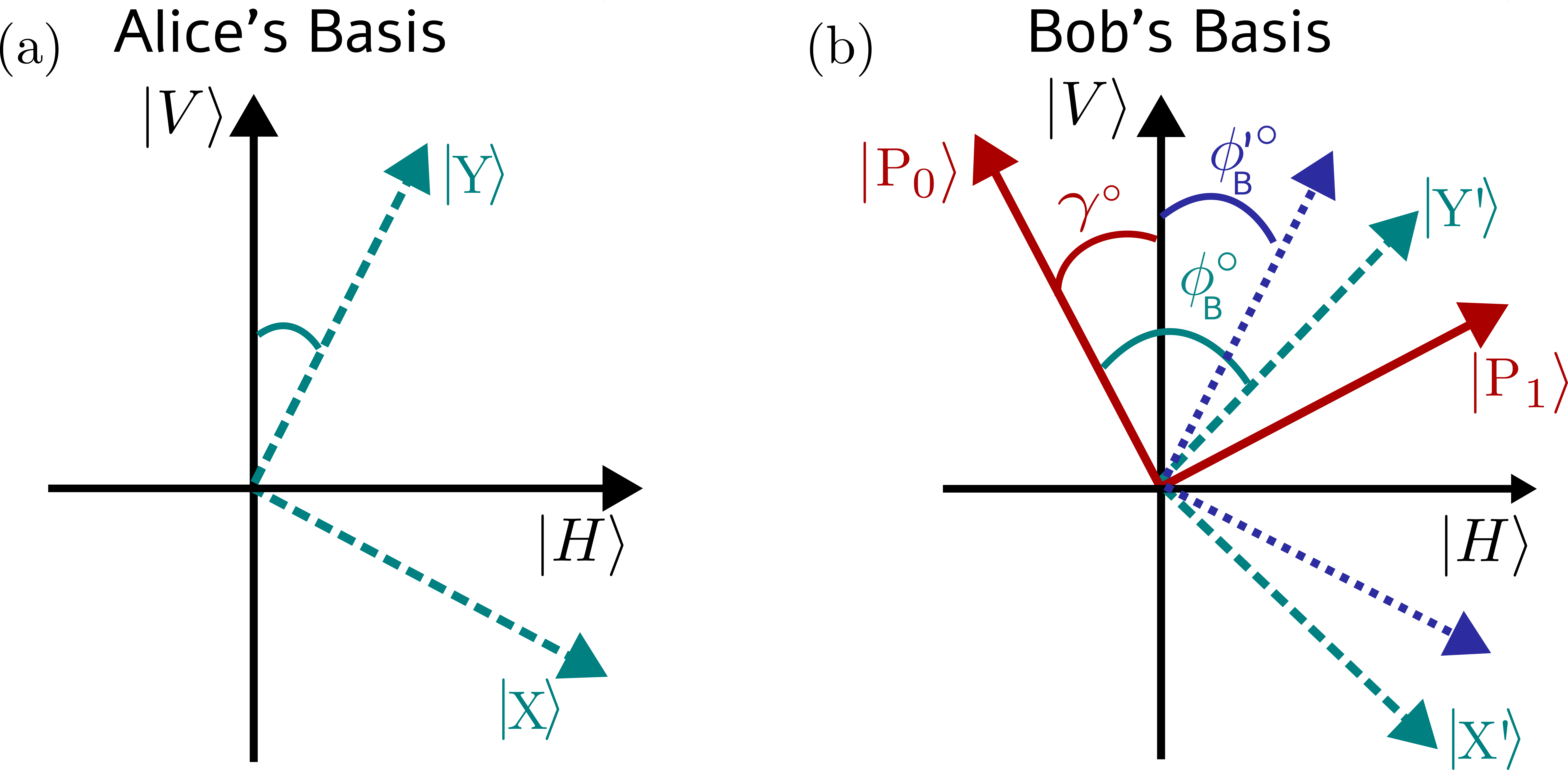}
    \caption{Polarization basis used by Alice and Bob for measuring the state $\rho_{AB}$. $\ket{E_0}$ and $\ket{E_1}$ are the PDL axis, rotated with an angle $\gamma$ with respect to the V and H axis. $\{\ket{X},\ket{Y}\}$ $(\{\ket{X'},\ket{Y'}\})$ represents the projection basis defined by an angle $\phi_A$ ($\phi_B$) for Alice and (Bob). Bob has to compensate for the basis rotation induced by the PDL by using the angle ($\phi_B'$). }
    \label{fig:basis}
\end{figure}
Here we explain the differences observed between the experimental results and theoretical predictions in Fig.~\ref{fig:Fritz} by considering a possible misalignment of the PDL axis. Let us start by writing the state produced by the AlGaAs source \eqref{State_Source} in the eigenbasis of the PDL $(P_0, P_1)$: 
\begin{equation}
\begin{aligned}
    \ket{\psi}=&\iint_{\omega_i \leq \omega_s} d\omega_sd\omega_iC(\omega_s,\omega_i)\ket{\omega_s,\omega_i}(\cos\gamma\ket{H,P_0}+\\&\sin\gamma\ket{H,P_1})+C(\omega_i,\omega_s)\ket{\omega_i,\omega_s}(\cos\gamma\ket{V,P_1}-\\&\sin\gamma\ket{V,P_0}) 
\end{aligned}
\end{equation}
where $\gamma$ is the angle of misalignment of the PDL eigenaxes with respect to the $(H,V)$ basis (see Fig.~\ref{fig:basis}) that can be controlled by the FPC placed just before the PDL emulator.
At the PDL output, the state becomes:
\begin{equation}\label{State_PDL_output2}
\begin{aligned}
    \ket{\psi}=&\frac{1}{\sqrt{N}}\iint_{\omega_i \leq \omega_s} d\omega_sd\omega_iC(\omega_s,\omega_i)\ket{\omega_s,\omega_i}\\ &[a\ket{H,P_0}+b\ket{H,P_1}+c\ket{V,P_1}+d\ket{V,P_0}]
\end{aligned}
\end{equation}
where $a=\sqrt{t_0}\cos\gamma$, $b=\sqrt{t_1}\sin\gamma$, $c=\sqrt{t_1}\cos\gamma$, and 
$d=-\sqrt{t_0}\sin\gamma$, with $t_0$ ($t_1$) the transmission coefficient of the polarization $P_0$ ($P_1$) and $N$ the normalization constant.
When the PDL axis are well aligned, i.e. $\gamma=0^\circ$ and $(P_0,P_1)=(V,H)$,
we retrieve Eq.~\eqref{partially_entangled_state}.

To evaluate the impact of this misalignment on the probability distribution $p(a_B,b_A,a_C,b_C)$, we write the state \eqref{State_PDL_output2} in an arbitrary basis $(X,Y)$, defined in Fig.~\ref{fig:basis}, representing the projection basis:
\begin{equation}
\begin{aligned}
    \ket{\psi}=&\frac{1}{\sqrt{N}}\iint_{\omega_i \leq \omega_s} d\omega_sd\omega_iC(\omega_s,\omega_i)\ket{\omega_s,\omega_i}\\ &[a'\ket{X,X'}+b'\ket{X,Y'}+c'\ket{Y,X'}+d'\ket{Y,Y'}]
\end{aligned}
\end{equation} where
\begin{widetext}
     
%\begin{equation}
%	\begin{aligned}
	%	a'&= a \sin\phi_A\cos\phi_B+b\cos\phi_A\cos\phi_B+c\cos\phi_A\sin\phi_B+d\sin\phi_A\sin\phi_B\\
		%b'&= -a\sin\phi_A\sin\phi_B-b\cos\phi_A\sin\phi_B+c\cos\phi_A\cos\phi_B+d\sin\phi_A\cos\phi_B\\
		%c'&= a\cos\phi_A\cos\phi_B-b\cos\phi_A\sin\phi_B+c\cos\phi_A\cos\phi_B+d\cos\phi_A\sin\phi_B\\
		%d'&=-a\cos\phi_A\sin\phi_B+b\sin\phi_A\sin\phi_B-c\sin\phi_A\cos\phi_B+d\cos\phi_A\cos\phi_B
	%\end{aligned}
%\end{equation}
%\end{widetext} 

\begin{equation}
    \begin{aligned}
		a'&= -a \cos\phi_A\sin\phi_B+b\cos\phi_A\cos\phi_B-c\sin\phi_A\cos\phi_B+d\sin\phi_A\sin\phi_B\\
		b'&= a\cos\phi_A\cos\phi_B+b\cos\phi_A\sin\phi_B-c\sin\phi_A\sin\phi_B-d\sin\phi_A\cos\phi_B\\
		c'&= -a\sin\phi_A\sin\phi_B-b\cos\phi_A\sin\phi_B+c\cos\phi_A\cos\phi_B-d\cos\phi_A\sin\phi_B\\
		d'&=-a\sin\phi_A\cos\phi_B+b\sin\phi_A\sin\phi_B+c\cos\phi_A\sin\phi_B+d\cos\phi_A\cos\phi_B
	\end{aligned}
\end{equation}
\end{widetext} 
$\phi_A$ ($\phi_B$) defines the basis used by Alice (Bob). To compensate for the rotation induced by the PDL when the incoming polarization is not parallel with $P_0$ or $P_1$, Bob uses the angle $\phi_B'$  defined by:
\begin{equation}
    \tan\phi_B=\tan(\phi_B'+\gamma)\sqrt{t_H/t_V}
\end{equation}
where the misalignment $\gamma$ is taken into account.
\begin{figure*}[t]
    \centering
    \includegraphics[scale=0.5]{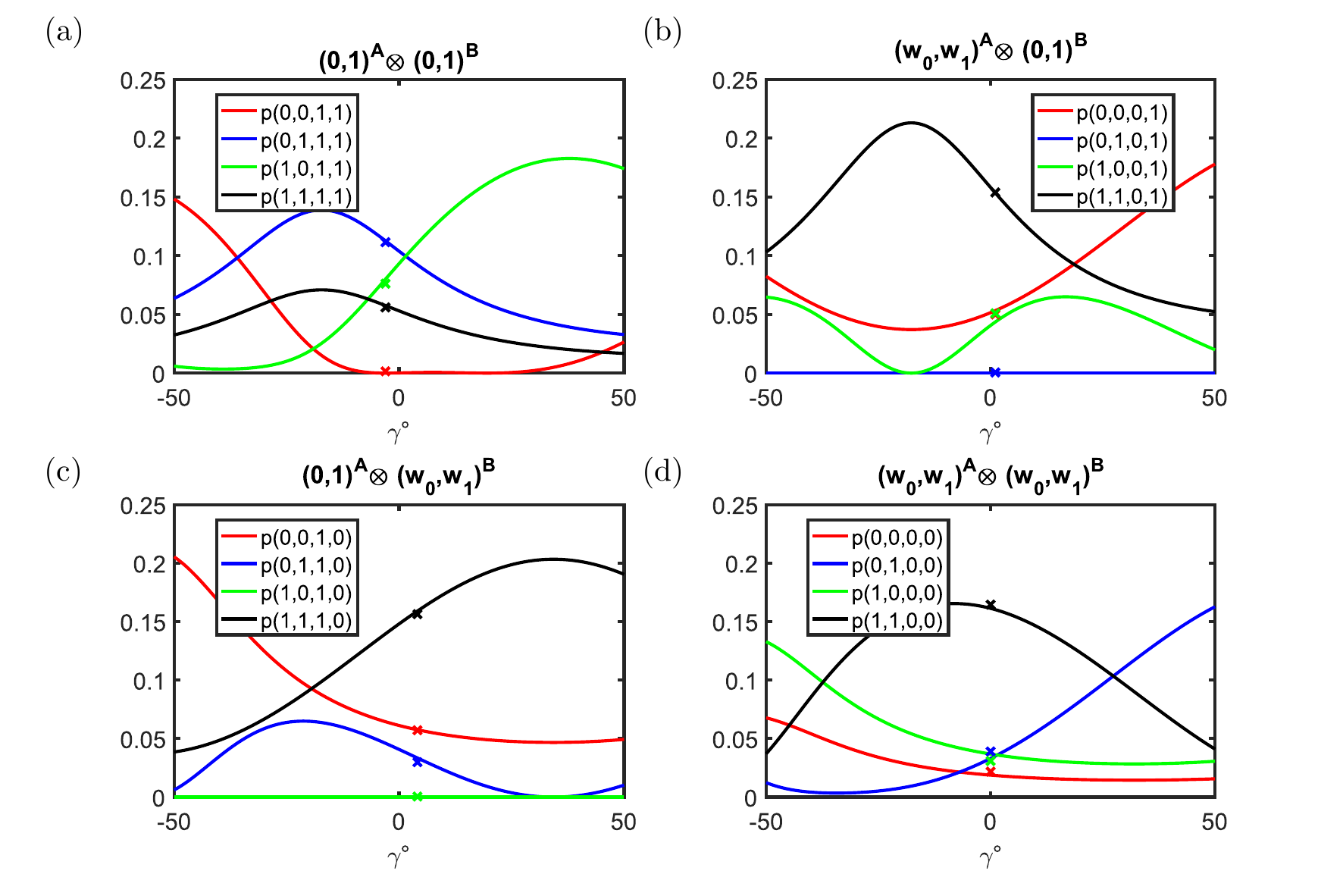}
    \caption{Probability of obtaining the different outcomes $p(a_B,b_A,a_C,b_C)$ as a function of the angle $\gamma$ related to the misalignment of the PDL axes with respect to the H and V axes. The crosses are the experimental data reported in Fig.~\ref{fig:Fritz}.}
    \label{fig:Prob}
\end{figure*}
\nocite{*}
The state $\rho_{AB}$ is projected into the $(0,1)$ or $(w_0,w_1)$ basis defined as follows:  
\begin{itemize}
    \item For $(0,1)$\hspace{0.57cm}: $\phi_A=-55.68^\circ$ , $\phi_B'=17.66^\circ$\\
    $\ket{X}=\ket{X'}=\ket{1}$ and $\ket{Y}=\ket{Y'}=\ket{0}$ 

    \item For $(w_0,w_1)$ : $\phi_A=-17.66^\circ$ , $\phi_B'=55.67^\circ$\\
    $\ket{X}=\ket{X'}=\ket{w_0}$ and $\ket{Y}=\ket{Y'}=\ket{w_1}^{A(B)}$ 
\end{itemize}
where we take as a reference $V=0^\circ$ and $H=90^\circ$.
The probabilities of obtaining the different outcomes are shown in Fig.~\ref{fig:Prob}. We clearly observe that the experimental data is well retrieved by varying $\gamma$ within a small range $[-3^\circ,4^\circ]$. Importantly, since $p(0,0,1,1)$, $p(0,1,0,1)$, and $p(1,0,1,0)$ don't change in this range, this possible small misalignment does not affect the value of $S_\Delta$.

% The \nocite command causes all entries in a bibliography to be printed out
% whether or not they are actually referenced in the text. This is appropriate
% for the sample file to show the different styles of references, but authors
% most likely will not want to use it.

\bibliography{Biblio}% Produces the bibliography via BibTeX.

\end{document}